\documentclass{article}

    \PassOptionsToPackage{numbers, compress}{natbib}
 \usepackage[preprint]{neurips_2025}
\usepackage[utf8]{inputenc} 
\usepackage[T1]{fontenc}

\usepackage{CJK}
\usepackage{array}      % 支持 p{width} 等
\usepackage{geometry} 
\usepackage[utf8]{inputenc} % allow utf-8 input
\usepackage[T1]{fontenc}    % use 8-bit T1 fonts
\usepackage{hyperref}    
\usepackage{url}            % simple URL typesetting
\usepackage{booktabs}       % professional-quality tables
\usepackage{amsfonts}       % blackboard math symbols
\usepackage{nicefrac}       % compact symbols for 1/2, etc.
\usepackage{microtype}      % microtypography
\usepackage{xcolor}         % colors
\usepackage{graphicx}
\usepackage{subfigure}
\hypersetup{breaklinks}
\usepackage{xspace}
\usepackage[linesnumbered,ruled,vlined]{algorithm2e}
\SetAlFnt{\normalfont}
\SetAlCapFnt{\normalfont}
\SetAlCapNameFnt{\normalfont}
\usepackage{enumitem}
\usepackage{multirow}
\usepackage{float}

\usepackage{bbm}%空心1

\usepackage{amsfonts}
\usepackage{bbm}
\usepackage{wrapfig}
\usepackage{ulem}
\usepackage{diagbox}
\usepackage{minitoc}
\usepackage{caption}
\usepackage{mathtools}
\usepackage{rotating}
\usepackage{listings}

\usepackage{cleveref}
\usepackage{amsmath,amsfonts,bm}

\def\eqref#1{equation~\ref{#1}}
\def\1{\bm{1}}

\DeclareMathAlphabet{\mathsfit}{\encodingdefault}{\sfdefault}{m}{sl}
\SetMathAlphabet{\mathsfit}{bold}{\encodingdefault}{\sfdefault}{bx}{n}

\newcommand{\me}{Auto-ATT\xspace}
\newcommand{\ac}{ATT-Corpus\xspace}
\newcommand{\atta}{ATT-Audios\xspace}
\DeclareUnicodeCharacter{2212}{-}

\title{Audio Turing Test: Benchmarking the Human-likeness of Large Language Model-based Text-to-Speech Systems in Chinese}

\author{%
  Xihuai Wang$^{1}$\thanks{Both authors contribute equally to this work.} \,\thanks{Work Done while interning at Meituan.},\, Ziyi Zhao$^{2}$\footnotemark[1] \,\thanks{Corresponding Author},\, Siyu Ren$^{2}$,\, Shao Zhang$^{1}$,\, \textbf{Song Li}$^{2}$,\, \textbf{Xiaoyu Li}$^{2}$,\and \textbf{Ziwen Wang}$^{2}$,\, \textbf{Lin Qiu}$^{2}$,\, \textbf{Guanglu Wan}$^{2}$,\, \textbf{Xuezhi Cao}$^{2}$,\, \textbf{Xunliang Cai}$^{2}$, \,\textbf{Weinan Zhang}$^{1}$ \\
  \\
  $^{1}$Shanghai Jiao Tong University, $^{2}$Meituan\\
  \texttt{zhaoziyi09@meituan.com} \\
}

\begin{document}

\maketitle

\begin{abstract}
Recent advances in large language models (LLMs) have significantly improved text-to-speech (TTS) systems, enhancing control over speech style, naturalness, and emotional expression, which brings TTS Systems closer to human-level performance.
Although the Mean Opinion Score (MOS) remains the standard for TTS System evaluation, it suffers from subjectivity, environmental inconsistencies, and limited interpretability. 
Existing evaluation datasets also lack a multi-dimensional design, often neglecting factors such as speaking styles, context diversity, and trap utterances, which is particularly evident in Chinese TTS evaluation.
To address these challenges, we introduce the \textbf{A}udio \textbf{T}uring \textbf{T}est (ATT), a multi-dimensional Chinese corpus dataset \ac paired with a simple, Turing-Test-inspired evaluation protocol. Instead of relying on complex MOS scales or direct model comparisons, ATT asks evaluators to judge whether a voice sounds human. This simplification reduces rating bias and improves evaluation robustness.
To further support rapid model development, we also finetune Qwen2-Audio-Instruct with human judgment data as \me for automatic evaluation. 
Experimental results show that ATT effectively differentiates models across specific capability dimensions using its multi-dimensional design. 
\me also demonstrates strong alignment with human evaluations, confirming its value as a fast and reliable assessment tool.
% We opensource the \ac and \me in Hugging Face \footnote[1]{The \ac and \me including the evaluated audio clips in this paper can be found in \url{https://huggingface.co/collections/meituan/audio-turing-test-682446320368164faeaf38a4}. }.
% The \ac and \me including the evaluated audio clips used in this paper can be found in \href{https://huggingface.co/collections/meituan/audio-turing-test-682446320368164faeaf38a4}{ATT Hugging Face Collection}.
% The white-box \ac and \me can be found in \href{https://huggingface.co/collections/AudioTuring/audio-turing-test-6826e24d2197bf91fae6d7f5}{ATT Hugging Face Collection}.% att version
The white-box \ac and \me can be found in \href{https://huggingface.co/collections/meituan/audio-turing-test-682446320368164faeaf38a4}{ATT Hugging Face Collection}.% meituan version

\end{abstract}

\section{Introduction}
% 第一段：语音合成拟人的重要性和挑战+由于模型的发展，针对传统TTS开发的评估数据集和方法难以区分LLM based方法的能力
% 第二段：具体来说在metrics方面 1）五点量表设计的缺陷例如缺乏可解释性导致eval的潜在biased，指标设计覆盖的维度过多无法准确识定位模型语音合成的能力问题 2）对humaneval人力的需求大且效率低，例如培训等问题 3）对modeleval而言？
% 第三段：数据集方面 1）缺少对不同的能力维度的考察？（对应我们数据构建方法覆盖的维度） 2）缺少陷阱数据验证带来的认知biased？（对应我们混入真人语音的陷阱设计）（这个部分需要进一步survey一下数据集）
% 第四段：我们需要构建一个新的方法，更好针对当前模型能力的痛点以及实现更高效低成本的评估迭代
% 第五段：针对metrics，ATT的改进
% 第六段：针对数据集，ATT的改进和构建方法
% 第七段：modeleval的实现和有效性+humaneval的结果
% 第八段：对metrics的有效性实验（对比MOS）

% \notes{From traditinal TTS to LLM-based TTS - Key Challenge for MOS in evaluation} 1) imoprtance of TTS especially in the , need evaluation 2) current metrics and dataset hard to differentiate the capability of models due to the fast development of LLMs
Text-to-Speech (TTS) technology plays a vital role in modern life by enhancing accessibility \cite{jain2025compact}, improving user interactions \cite{wang2024contextual}, and supporting diverse applications such as voice assistants \cite{10.1145/3659625} and audiobooks \cite{yeh2024dialog}.
The advancement of large language models (LLMs) technology brings TTS a significant breakthrough \cite{anastassiou2024seed}.
The improved controllability of speech style and intonation brought by LLMs \cite{li2024styletts}, along with enhancements in speech naturalness and emotional expression \cite{wang2025spark}, has transformed models from merely approaching human-like performance to rivaling human speech. 
Such technological advancements have introduced new challenges to distinguish the performance of the latest LLM-based TTS systems.

Current TTS benchmarks still rely on human listening tests as a golden metric, typically the 5-point Mean Opinion Score (MOS) \cite{ITU-P808} and its variants like Comparative MOS (CMOS). 
However, MOS has well-known flaws: scores drift with listeners, playback setups, and even day-to-day moods \cite{chiang23_interspeech}; controlling this variance requires expensive rater training and calibration, and repeated studies often yield conflicting results for the same system \cite{kirkland2023stuck}.  
Attempts to replace listeners with MOS-prediction models like UTMOS \cite{saeki2022utmos} and DMSMOS \cite{reddy2022dnsmos} falter because those models rarely transfer across corpora \cite{perrotin23_blizzard}. 
% Concurrently, the enhanced control over speech style and intonation enabled by LLMs \cite{anastassiou2024seed}, coupled with improvements in speech naturalness and emotional expressiveness, has elevated these models from merely approximating human-like qualities to genuinely competing with actual human speech.
Given that state-of-the-art LLM-based TTS systems perform comparably at the upper end of the MOS scale \cite{LEMAGUER2024101577}, MOS is increasingly recognized as both noisy and too coarse-grained to meaningfully distinguish between current high-quality TTS systems.

Beyond the limitation of MOS, current corpus data used in TTS evaluation typically serve general purposes across multiple tasks or training scenarios \cite{anastassiou2024seed,wang2025spark}, lacking a specialized design to evaluate multidimensional capabilities. 
For instance, standardized evaluations often prioritize overall speech quality in limited contexts, neglecting diverse speaking styles, varied linguistic scenarios, or specialized content.
Additionally, listening tests rarely incorporate hidden human speech references or intentionally crafted trap utterances, undermining their ability to effectively identify evaluator biases and attention distribution \cite{chiang23_interspeech}. These limitations become particularly pronounced in the Chinese linguistic context, where factors like prosodic pauses \cite{lavin2002issues}, multilingual code-switching \cite{yang2024bilingual}, polyphonic characters \cite{dai2025disambiguation}, and special symbols significantly influence speech fluency and naturalness.
Consequently, the absence of multidimensional and trap data in existing evaluation datasets \cite{chiang23_interspeech} magnifies the inherent shortcomings of MOS-based assessments, restricting the ability of current TTS evaluations to yield comprehensive and discriminative insights.

% Moreover, due to the continuous advancement of model capabilities, current evaluation datasets also face issues related to insufficient coverage.

\begin{figure}
\vspace{-10pt}
    \centering
    \includegraphics[width=0.95\linewidth]{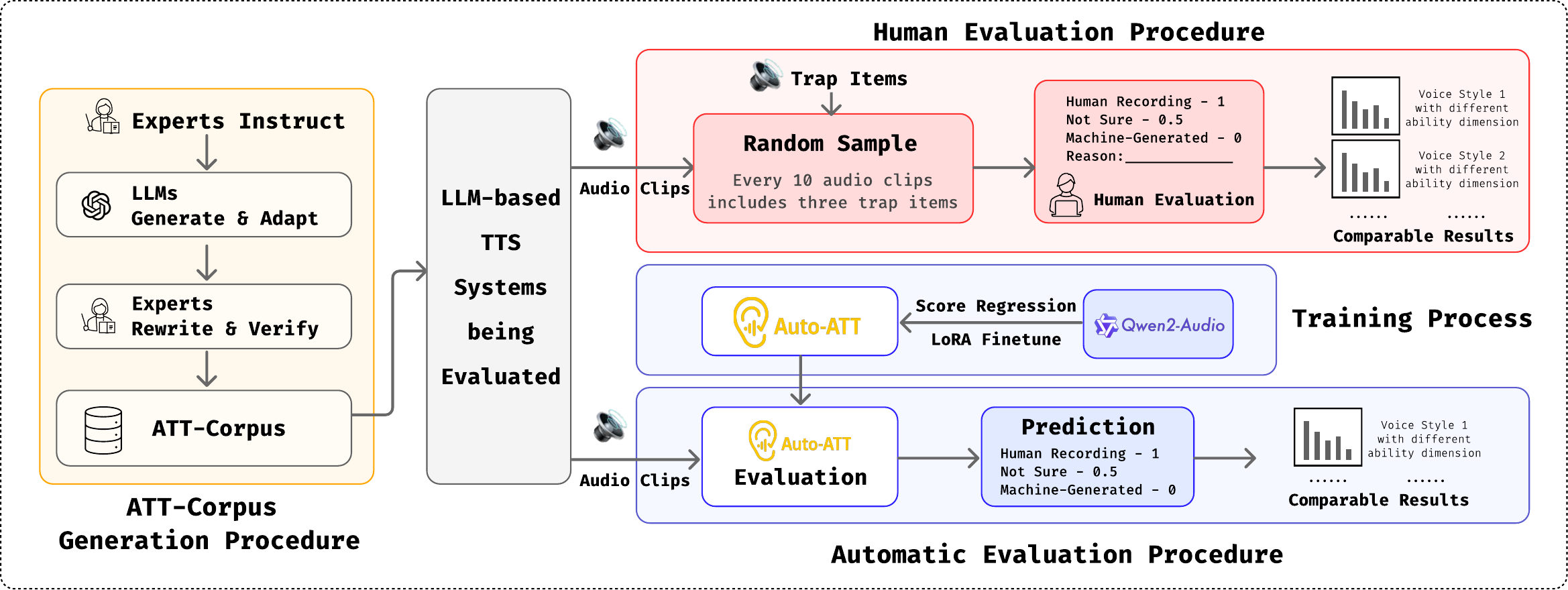}
    \caption{\textbf{Audio Turing Test Evaluation Framework}: (1) Corpus Generation: a semi-automatic corpus generation pipeline for generating the challenge TTS synthesis corpus for ATT evaluation; (2) Human Evaluation: a human-evaluation protocol that enables precise, comparable assessments and lowers evaluation costs through a simple yet effective Turing-test-style design, (3) Automatic Evaluation: \me, an automatic tool to predict the Human-likeness Score for rapid iterations. }
    \label{fig:protocol}
    \vspace{-20pt}
\end{figure}

% \notes{We need a clear and effective evaluation method for LLM in evaluation.}
% To better evaluate advanced LLM-based TTS methods in Chinese, there is an urgent need for a more comprehensive, efficient, and discriminative approach.
Inspired by the classic Turing Test \cite{french2000turing}, as shown in \Cref{fig:protocol}, we propose the \textbf{A}udio \textbf{T}uring \textbf{T}est (ATT), an evaluation framework combining a multi-dimensional dataset \ac with a Turing test-based evaluation protocol and metrics.
To evaluate the human-likeness of Chinese TTS systems, we first built a targeted evaluation corpus addressing key challenges in Chinese speech synthesis.
Based on the \ac, we design a simple and easy-implement human evaluation protocol.
By requiring evaluators to provide ternary judgments on whether each sample is human, along with brief justifications, ATT facilitates both quantitative and qualitative assessments of speech human-likeness. 
This approach mitigates the anchoring effects and lack of cross-context comparability commonly associated with traditional scale-based methods such as MOS.
ATT employs randomized clip assignment, trap items for attention monitoring, and expert-validated justifications to ensure data quality, supporting reliable, unbiased clip-level analysis.
To enable swift automated evaluation and accelerate TTS model iteration, we fine-tuned Qwen2-Audio-Instruct~\cite{chu2024qwen2} on a rigorously annotated ATT dataset, producing \me.

Using the ATT protocol, we collected ratings from 857 native Chinese listeners through crowdsourcing platforms.
Experimental results demonstrate that ATT is a sharp and reliable evaluation framework.
Benchmarking results further indicate that ATT effectively distinguishes the performance of different TTS models.
Notably, even the top-performing model, Seed-TTS \cite{anastassiou2024seed}, achieves only a human-likeness score of 0.4 on ATT—considerably lower than that of real human speech, and in stark contrast to previously reported MOS scores.
Analyses across sub-dimensions and voice styles demonstrate that ATT enables multi-axis evaluation of LLM-based TTS systems and supports direct cross-system comparisons.
We assess the effectiveness of \me through trap item tests and comparing auto-evaluation results with human ratings.
\me significantly outperforms traditional MOS predictors in evaluating trap clips and shows strong alignment with human scores.
% This order held across all five skill areas: Seed-TTS led in code switching and numeral handling and fell behind only on classical texts. The results confirm that ATT both measures performance precisely and reveals the specific flaws that listeners notice.
% \notes{\me and its effectiveness with humaneval}

% \notes{experiment for metric effectiveness}

In summary, our contributions are as follows: 
\begin{itemize}
    \item We introduce the Audio Turing Test, an evaluation framework comprising a multi-dimensional Chinese corpus (\ac) and a Turing Test-inspired protocol, designed to effectively assess the human-likeness of LLM-based TTS systems.
    \item We further train \me on human evaluation data to develop an automatic evaluation tool that enables fast and effective assessment of TTS systems, demonstrating its effectiveness through strong consistency with human ratings.
    \item We benchmark state-of-the-art LLM-based TTS systems using both quantitative and qualitative analyses, thereby validating the effectiveness and robustness of the ATT framework.
\end{itemize}

\section{Related Works}

The quality of Text-to-Speech (TTS) systems is typically assessed through a combination of objective metrics and subjective human evaluations.

Objective metrics focus on quantitative comparisons between synthesized speech and reference speech. 
Speaker similarity (SIM) is a commonly used objective metric, widely adopted in recent research involving LLM-based TTS systems \cite{Wang2023NCLM,anastassiou2024seed}. 
However, since SIM requires comparison to reference speech, only the model trainer can use SIM as an evaluation metric, which is challenging to perform horizontal comparisons across different TTS systems.
Moreover, speaker similarity only indicates how closely the voice matches the reference speaker; it does not capture any other aspects of choral quality \cite{guner2012analysis}.
In addition, some prediction systems based on human evaluation results, such as UTMOS \cite{saeki2022utmos} and DMSMOS \cite{reddy2022dnsmos}, have been used to assess the quality of speech generated by TTS systems. 
However, these supervised scoring models often suffer from generalization issues, making it difficult to evaluate new TTS systems reliably.

Currently, the majority of TTS systems' research continues to use Mean Opinion Score (MOS) \cite{streijl2016mean} listening tests as the ``gold standard,'' commonly employing the five-level absolute category rating scale and its derivatives such as CMOS, CCR \cite{Naderi2021CCR}, and MUSHRA-style multi-stimulus methodologies \cite{ITU-R-BS1534-3}.
Recent studies have systematically exposed several structural flaws inherent in MOS style evaluations. 
Due to the subjective nature of MOS, inconsistencies in scales or instructions used across experiments render scores of identical speech samples incomparable between studies \cite{kirkland2023stuck}. 
Additionally, listeners' prior assumptions about the testing task or usage scenario significantly affect scoring distributions \cite{edlund2024assessing}.
For the comparative methods like CMOS, the presence of lower-quality TTS systems may affect the MOS scores of higher-quality TTS systems \cite{LEMAGUER2024101577}.
Comparison-based methods such as MUSHRA \cite{ITU-R-BS1534-3} require listeners to compare synthesized speech with a human reference, which may introduce scoring bias due to the presence of that reference \cite{varadhan2024rethinking}.
CMOS may also suffer from limited discriminatory power, especially when test items are of similar perceived quality, making it challenging to distinguish meaningful differences.
To mitigate these issues, some researchers have incorporated relative pairwise comparisons and grouping analyses in naturalness assessments, reporting enhanced sensitivity in distinguishing between systems \cite{perrotin23_blizzard}.

% Despite these efforts, subjective evaluations still face unresolved fundamental challenges. Contemporary state-of-the-art TTS systems have surpassed the discriminative capabilities of traditional scales. 
% Achieving both fine-grained and cost-effective human annotations while maintaining cross-study comparability remains an unresolved issue. 
% Furthermore, consensus is lacking regarding multilingual scale consistency, controlling listener cultural differences, and the reliability of automated alternative metrics. 
% These unresolved challenges indicate that MOS and its current improvements are inadequate for evaluating next-generation "human-like" speech synthesis systems.

Although evaluation-focused studies emphasize that TTS research should fully document experimental procedures and provide standardized templates, current TTS experiments frequently lack detailed reporting on human listening test procedures, such as listener screening, compensation, and interface instructions, making replication difficult \cite{chiang23_interspeech}.
Moreover, with the advent of LLM-related technologies, the synthesized speech quality of new TTS systems approaches that of human speech, leading to a ceiling effect and insufficient resolution in MOS, unable to discern subtle differences among high-end systems \cite{LEMAGUER2024101577}. 
Consequently, there is an urgent need for a novel evaluation methodology, complete with a comprehensive implementation protocol and accompanying multidimensional test datasets, to enable more thorough and reliable subjective evaluations of TTS systems.

\section{Audio Turing Test}\label{sec:att}

To address the challenges in the current subjective evaluation of TTS systems, we design the Audio Turing Test (ATT). 
ATT is an evaluation framework with a standardized human evaluation protocol and an accompanying dataset \ac, aiming to resolve the lack of unified protocols in TTS evaluation and the difficulty in comparing multiple TTS systems. 
Moreover, for comprehensive evaluation, \ac is designed with appropriate dimensions to help identify specific capability differences among TTS systems. 
To further support the training and iteration of TTS systems, we utilized additional private evaluation data to train \me based on Qwen2-Audio-Instruct via LoRA~\cite{hu2022lora} finetuning, enabling a model-as-a-judge approach for rapid evaluation of TTS systems on the \ac.
In this section, we provide a detailed description ofof the construction of the \ac, ATT evaluation protocol design along with the \me.

\subsection{\ac Dataset}\label{sec:dataset}

Currently, TTS evaluation primarily relies on a subset of samples selected from publicly available speech datasets. 
This results in limited coverage and makes assessing a model's ability to synthesize complex speech challenging.
We construct \ac as a comprehensive corpus for TTS evaluation to address this limitation.
Taking Chinese as a representative example, we first identify the key challenges TTS systems face, which guide the two-stage data production process of \ac.

\paragraph{Data Description.}
We categorize the linguistic capabilities required for Chinese TTS synthesis based on the linguistic phenomena in the corpus to construct a dataset tailored for ATT evaluation.
The corpus covers five key dimensions of Chinese linguistic competence: (1) Special Characters and Numerals, (2) Chinese-English Code-switching, (3) Paralinguistic Features and Emotions, (4) Classical Chinese Poetry/Prose, and (5) Polyphonic Characters.
The detailed composition of the corpus is presented in \Cref{tab:cropus}.

\begin{table}[]
\vspace{-10pt}
    \centering
        \setlength{\tabcolsep}{3.5pt} 
        \caption{\textbf{Corpus Examples of \ac.}}
        % \resizebox{\linewidth}{!}{
        {\scriptsize
    \begin{tabular}{
>{\raggedright\arraybackslash}p{2cm}
>{\raggedright\arraybackslash}p{5cm}
>{\raggedright\arraybackslash}p{6cm}}
    \toprule
       \textbf{Dimension}  & \textbf{Description} & \textbf{Example} \\
       \midrule
       Special Characters and Numerals  & Analyze the numbers, special characters, letters, and other information types in the text and transcribe them into the most appropriate or commonly used pronunciations. &\begin{CJK}{UTF8}{gkai} 我们公司也有些年头了呢。 \uline{2010 年 6 月 8 日}的时候公司刚成立，现在算算已经快满\uline{ 12 年}了，真的是时间过得挺快的。这一路走来也不容易啊。 \end{CJK}\\
       \midrule
        Chinese-English Code-switching  & Primarily Chinese, interspersed with a few words from other languages, used to assess whether the pronunciation is accurate. & \begin{CJK}{UTF8}{gkai}没想到\uline{B站}有这么多不同类型的片子，昨晚我\uline{在 bilibili 上}看了一部新的纪录片......\end{CJK} \\

       \midrule
       Paralinguistic Features and Emotions  & Expressive paralinguistic phenomena, such as laughter, and the expression of various emotional states. & \begin{CJK}{UTF8}{gkai}\uline{呜呼，终于下班了}。今天的工作简直让人崩溃，真是忙得一刻都没停过。溜了溜了，赶紧回家休息了，我感觉一回家就要睡着，等会晚点去个洗脚城好好放松一下。 \end{CJK}\\
       \midrule
       Classical Chinese Poetry/Prose  & Each character in classical Chinese poetry and prose is pronounced correctly in terms of its initial consonant, final, tone, and other aspects of articulation. & \begin{CJK}{UTF8}{gkai}苏辙笔下长江的描绘：\uline{“出西陵，始得平地，其流奔放肆大。”}江水奔腾不息、气势磅礴的景象让人震撼不已。三峡之行...... \end{CJK}\\
               \midrule
       Polyphonic Characters  & Polyphonic Chinese characters are pronounced correctly. & \begin{CJK}{UTF8}{gkai} 老中医说，这病\uline{症}得慢慢调理，着急不得。可这病的\uline{症}结到底在哪呢？\end{CJK}\\
         \bottomrule
    \end{tabular}
    }
    \vspace{-15pt}
    \label{tab:cropus}
\end{table}

\paragraph{Corpus Generation and Verification.}
To reduce manual labor costs and ensure the long-term sustainability of the corpus production process, we adopt a semi-automated approach that combines initial generation and adaptation using large language models (LLMs), followed by expert revision and validation.
We employ GPT-4o \cite{hurst2024gpt} as the primary model for initial corpus generation.
We generate base corpora across various linguistic categories using the prompt and sample text illustrated in the figure.
Subsequently, we utilize DeepSeek-R1 \cite{guo2025deepseek} to perform colloquial adaptation in Chinese, enhancing the naturalness and human-likeness of the generated text.
After the automated generation process, four linguistics experts\footnote{Experts refer to individuals holding a master's degree in linguistics or a related field.} conducted standardized revisions of the corpus. 
The prompts for data generation, along with the specific revision and review guidelines, are provided in \Cref{app:ATTdata}.
Upon completion of the revisions, the experts conducted cross-checking to ensure the quality of the corpus.

\paragraph{Black-box and White-box.}
To ensure a fair and reliable evaluation, we divide the generated data into white-box and black-box subsets.
The white-box subset is made publicly available, while the black-box subset is hosted on an evaluation platform for open and blind testing\footnote{The black-box data will be hosted on the \href{https://agi-eval.cn/mvp/home }{AGI-Eval} Platform, where both human and automatic evaluations will be conducted.}.
Our experiments validate the consistency between white-box and black-box evaluation results.

\subsection{Evaluated Audio Clips Generation and Validation}
After completing the corpus collection, we generate audio clips using the TTS models to be evaluated.
To ensure evaluation accuracy, we perform manual spot checks on the synthesized speech with the involvement of two expert reviewers.
To balance the sample's representativeness with the efficient use of human review resources, a sampling rate of 25\% is adopted. 
Specifically, we examine two aspects during this stage: synthesis success and synthesis consistency. 
The details of the validation process are in \Cref{app:attaudio}.
Note that at this stage, we do not evaluate or inspect the human-likeness of the synthesized speech.

% \textbf{Synthesis Success. }Synthesis success refers to the correctness of the output audio in terms of overall audio quality, transcription accuracy, and language appropriateness. 
% Specifically, we check for issues such as: significant audio quality defects (e.g., excessive robotic noise, jittering), extremely short or incomplete audio that cannot be properly transcribed (e.g., only a single “ah” sound or complete silence), language mismatch (e.g., input in Chinese but output in Japanese), inconsistencies in voice timbre within a single clip (e.g., mixing multiple voice styles), and other cases where the output is unintelligible in the target language.

% \textbf{Synthesis Consistency. }Synthesis consistency refers to the consistency of output when the same text is synthesized multiple times using the same voice and technology. This assessment focuses on whether the resulting audio clips are consistent in overall characteristics such as voice timbre (e.g., gender, age), language (e.g., remaining within the same language such as Chinese or English), and prosody (e.g., intonation, stress, and tone of voice). The goal is to determine whether the outputs can reliably be attributed to the same voice.

\subsection{Human Evaluation Protocol}\label{sec:protocol}

In the ATT human evaluation, participants completed a forced-choice speech-authenticity test. As shownshown in \Cref{fig:protocol}, we propose the following protocol to implement ATT:

% Every participant evaluate the random excerpts that were drawn without replacement from a larger pool across models. 
% To monitor attentiveness and protect data integrity, in each 10 excerpts during the process for one participant, there are the three trap items which are one deliberately flawed synthetic clip (e.g., \todo{}) and two bona-fide human recordings. 
% After listening to each audio excerpt, they selected one of three mutually exclusive labels: [Human Recording], [Machine-Generated], or [Not Sure]. 
% Then the participant enter a short free-text justification for their decision for supporting qualitative error analysis. 
% Due to inattentive respondents tend to misclassify such clearly distinguishable stimuli, responses are deemed valid only if the participant correctly flags the flawed synthetic item and correctly identifies at least one of the two human clips in each 10 excerpts. 
% All data from participants who failed this attention check were excluded from subsequent analyses, thereby ensuring that the retained corpus reflected judgments made with minimal guessing or fatigue.
% In addition, after the data is collected, expert reviewers manually evaluate the consistency between participants' free-text justification and their annotated labels. 
% Any data that fails the consistency check will also be excluded.

\textbf{Sampling and Assignment.} Each participant is randomly assigned audio clips sampled without replacement from a pool containing outputs from multiple TTS models.

\textbf{Attention Monitoring via Trap Items.} To ensure participant attentiveness, we include trap items at regular intervals. Specifically, three trap items are used for every 10 clips: one deliberately flawed synthetic clip and two genuine human recordings. We also open source these trap items in \atta for future evaluation.

\textbf{Labeling and Justification.} For each audio clip, participants select one of three labels: [Human], [Unclear], or [Machine]. They also required to provide a short free-text justification to support subsequent qualitative analysis.

\textbf{Attention Check Validation.} The response batch of participants is considered valid only if they correctly identify the deliberately flawed synthetic clip and at least one of the two human recordings within each 10-clip set. Responses that fail to meet this criterion are excluded from further analysis.

\textbf{Expert Consistency Review. }After data collection, expert reviewers assess whether participants' free-text justifications align with their labels. Any responses that fail this consistency check are also excluded from analysis.

Each excerpt and its corresponding judgment were treated as an independent sampling unit in our protocol design. The random assignment of excerpts, combined with the absence of inter-item feedback, minimized learning effects and reduced inter-trial dependence, enabling clip-level modeling of classification accuracy.

To validate the protocol's effectiveness, we report results from a mixed-effects logistic regression analysis, with participants modeled as a random effect, using a generalized linear mixed model (GLMM) \cite{bolker2009generalized}.
% And such a design enables fine-grained comparisons across synthesis methods or linguistic contexts that would be obscured in participant-level aggregates. 

\subsection{Human-likeness Score}

% Based on the evaluation protocol, we define the metric in ATT, Human-likeness Score (HLS).

% There is one human label of audio $i$ that collected in $\mathcal{L}$:
% \[
% \mathcal{L}=\{[\text{Human Recording}], [\text{Not Sure}], or [\text{Machine-Generated}]\}
% \]
% % For each audio clip, we define its HLS score as:
% % \begin{equation}
% %     \textbf{HLS}_n = 
% %     \begin{cases}
% %     1, & \text{Human Label = Human Recording}\\
% %     0.5,   & \text{Human Label = Not Sure}\\
% %     0, & \text{Human Label = Machine-Generated}
% % \end{cases}
% % \end{equation}

% The score of audio \(i\) is then the fraction of satisfied labels:
% \[
% s_{i}=
%   \begin{cases}
%     1,& \text{if audio }i\text{ satisfies label }\text{[Human Recording]},\\
%     0.5, & \text{if audio }i\text{ satisfies label }\text{[Not Sure]},\\
%     0, & \text{if audio }i\text{ satisfies label }\text{[Machine-Generated]}
%   \end{cases}
% \]

% \paragraph{Model-level metric (HLS)}%
% Given \(N\) audio clips produced by the model,
% \(\mathcal{S}=\{s_1,\dots,s_N\}\),
% the model's HLS is defined as
% \[
% \mathrm{HLS} :=
%   \frac{1}{N}\sum_{i=1}^{N} s_i.
% \tag{2}
% \]

Based on the evaluation protocol, we define a metric to quantify the human-likeness of audio clips synthesized by TTS systems: the Human-likeness Score (HLS).

There is one human label for each audio clip $i$ collected in the set $\mathcal{L}$:
$$
\mathcal{L} = \{\text{Human}, \text{Unclear}, \text{Machine}\}
$$

In HLS, the individual scores for each audio clip $i$ are then expressed using the indicator function $\mathbbm{1}({\cdot})$:
$$
s_i = \mathbbm{1}(\text{Label} = \text{Human}) + 0.5 \cdot \mathbbm{1}(\text{Label} = \text{Unclear})
$$
% $$
% s_i =
%   \begin{cases}
%     1, & \text{if the label is } \text{Human}, \\
%     0.5, & \text{if the label is } \text{Not Sure}, \\
%     0, & \text{if the label is } \text{Machine-Generated}.
%   \end{cases}
% $$
% In S-HLS, the score for audio clip $i$ is defined as follows:
% $$
% s_i^{\text{strict}} = \mathbbm{1}\{\text{Label} = \text{Human}\}
% $$

% $$
% s_i =
%   \begin{cases}
%     1, & \text{if the label is } \text{Human Recording}, \\
%     0, & \text{otherwise}.
%   \end{cases}
% $$

Given $N$ audio clips produced by one TTS system, represented as the set $\mathcal{S} = \{s_1, \dots, s_N\}$, the system's HLS is defined as the average of the individual scores $s_i$:
% Given $N$ audio clips produced by one TTS system, represented as the set $\mathcal{S} = \{s_1, \dots, s_N\}$, the system's HLS is defined as the average of the individual scores $s_i$ and $s_i^{\text{strict}}$:
% $$
% \mathrm{HLS} = \frac{1}{N} \sum_{i=1}^{N} s_i, \text{and } \mathrm{S}\text{-}\mathrm{HLS} = \frac{1}{N} \sum_{i=1}^{N} s_i^{\text{strict}}.
% $$

$$
\mathrm{HLS} = \frac{1}{N} \sum_{i=1}^{N} s_i~.
$$

% S-HLS provides a more stringent evaluation criterion by exclusively counting audio clips labeled explicitly as ``Human,'' excluding ambiguous cases. 
We employ HLS to quantify the human-likeness of a TTS system's speech synthesis, which can be assessed both overall and within specific sub-dimensions.
The resulting numeric HLS scores can also supervise the training of automated prediction models.

\subsection{\me}\label{sec:train}

To facilitate rapid evaluation iterations and enhance the usability of the assessment process, we fine-tuned Qwen2-Audio-Instruct\footnote{\url{https://github.com/QwenLM/Qwen2-Audio}, Apache 2.0 License} \cite{chu2024qwen2} on a subset of human evaluation data to enable a ``model-as-a-judge'' approach that allows the model to predict Human-likeness Score (HLS).

\paragraph{Data.} For training \me, we collect both positive and negative samples of ATT HLS through multiple rounds of human annotation on audio data generated by four model families, focusing on three capability subsets from the ATT dataset. 
The training subsets include Chinese-English code-switching, character-level pronunciation, paralinguistics, and emotion.
We used internal TTS systems to synthesize the speech via a corpus of the subsets, and recruited 437 annotators from crowdsourcing platforms to evaluate the audio clips following our protocol. 
Three annotators labeled each audio clip to arrive at a final label.
Within each model family, we reserved one voice for the test set and used the remaining 3 voices for training. 
During training, each batch of audio samples was drawn from a single subset to maintain consistency.

% Data collection details can be found in \Cref{app:traindata}.

\paragraph{Training.}
We utilized TTS-generated speech segments accompanied by instructional prompts designed to guide the model in evaluating speech human-likeness. These inputs were employed to adapt Qwen2-Audio-Instruct for HLS prediction.

Though originally introduced as an auto-regressive audio language model, we adapt Qwen2-Audio-Instruct for ATT score regression by leveraging the logits from its existing \texttt{lm\_head}. Specifically, we selected three semantically significant tokens: Human, Unclear, and Machine, whose logits represent the model's internal judgments regarding speech quality. A Softmax function was applied to these logits to obtain a normalized probability distribution across the three quality categories. Subsequently, this distribution was converted into a weighted average score by associating each category with a predefined discrete ATT score value: 1 for Human, 0.5 for Unclear, and 0 for Machine. The predicted HLS was calculated as follows:

\begin{equation}
s_{i}^{\text{pred}} = \sum_{\text{Label}} P(\text{Label})\cdot \left[1\cdot\mathbbm{1}(\text{Label}=\text{Human}) + 0.5\cdot\mathbbm{1}(\text{Label}=\text{Unclear})\right]
\end{equation}

% \begin{equation}
% \text{HLS}_{\text{pred}} = \sum_{\text{Label}} \text{Prob}(\text{Label}) \times \text{Score}(\text{Label}),
% \end{equation}

% where

% \begin{equation}
% \text{Score}(\text{Label}) =
% \begin{cases}
% 1, & \text{if Label = Human} \\
% 0.5, & \text{if Label = Not Sure} \\
% 0, & \text{if Label = Machine-Generated}
% \end{cases}
% \end{equation}

Logits were specifically extracted from the final token position of the input prompt, denoted by the character ``:''. The input prompt comprises both audio content and instructional guidance.

During training, we adopted a loss function consisting of a weighted linear combination of Mean Squared Error (MSE) and Bradley-Terry (BT) \cite{hunter2004mm} losses:

\begin{equation}
\mathcal{L}_{\text{Total}} = 0.4 \times \mathcal{L}_{\text{BT}} + 0.6 \times \mathcal{L}_{\text{MSE}}~,
\end{equation}
where $\mathcal{L}_{\text{BT}} = -\sum_{(i,j), s.t., s_{i}^{gt} > s_{j}^{gt}}
\log\sigma\left(s_{i}^{\text{pred}} - s_{j}^{\text{pred}}\right)$ and $\mathcal{L}_{\text{MSE}} = \frac{1}{2}\sum_{i}\left(s_{i}^{\text{pred}} - s_{i}^{\text{gt}}\right)^2$.

The model fine-tuning employed Low-Rank Adaptation (LoRA) with hyperparameters configured as follows: rank ($r$) of 32, scaling factor ($\alpha$) of 32, and dropout rate of 0.05. LoRA adapters were applied exclusively to all linear layers within the LLM component of Qwen2-Audio-Instruct, while other parameters remained fixed throughout the training process.
We used 4 NVIDIA A100 GPUs to train \me, which takes about 1 hour. The server's CPU was an Intel Xeon Platinum 8358P (2.60 GHz, 128 cores).

\section{Experiments}\label{sec:exp}

The evaluation involves a total of 20 voice styles across 5 TTS model families, as detailed in \Cref{tab:models}.
% All the audio clips are using API to generate, in which a total of 23,754 audio clips.
% We also opensource these generated audio clips in \atta as a baseline for future evaluation.
\begin{table}[]
    \centering
        \caption{\textbf{The model families and their voice styles we evaluated, introducing in \Cref{app:tts}}.}\label{tab:models}
    \resizebox{\linewidth}{!}{
    \begin{tabular}{ll}
    \toprule
      \textbf{Model Families}   & \textbf{Voice Styles}  \\
      \midrule
        Cosyvoice2.0 \cite{du2024cosyvoice} & longshuo, longxiaocheng, longxiaochun, longxiaoxia\\
        \midrule
        MiniMax-Speech \cite{minimax_speech01} & Male\_botong\_platform, Podcast\_girl\_platform, Boyan\_new\_platform, Female\_yaoyao\_platform\\
        \midrule
        \multirow{2}{*}{Seed-TTS \cite{anastassiou2024seed}} & Sky (zh\_female\_shuangkuaisisi\_moon\_bigtt), Alvin (zh\_male\_wennuanahu\_moon\_bigtts), \\
        &Brayan (zh\_male\_shaonianzixin\_moon\_bigtts), Moon (zh\_female\_linjianvhai\_moon\_bigtts)\\
        % Spark-TTS \cite{wang2025spark} & x4\_lingfeizhe\_oral, x4\_lingxiaoxuan\_oral, x4\_lingyuzhao\_oral\\
        \midrule
        Step-Audio \cite{huang2025step} & qingniandaxuesheng, shenchennanyin, linjiajiejie, wenjingxuejie\\
        \midrule
        GPT-4o \cite{hurst2024gpt} & Alloy, Shimmer, Echo, Onyx\\
        \bottomrule
    \end{tabular}
    }
    \vspace{-15pt}
\end{table}

\subsection{Human Evaluation}
% \todo{new data for 2.2}
Following the ATT human evaluation protocol outlined in \Cref{sec:att}, we recruited 857 native Chinese speakers through crowdsourcing to evaluate the TTS systems. The participant pool included 202 males, 247 females, and 408 who select `Prefer not to say.'
% The detail demographics of the participants can be found in \Cref{}.
As shown in \Cref{fig:ui}, in each evaluation phase, participants will listen to an audio clip and make a single-choice selection afterward, choosing whether the source of audio is [Human] - 1, [Unclear] - 0.5, or [Machine] - 0. 
Participants were further required to provide written justifications for each of their judgments, which supports a deeper qualitative analysis of the perceptual and decision-making processes underlying their evaluations.
Each audio clip took approximately 45 seconds to 1 minute to evaluate and annotate.
Compensation was provided at a rate of 0.8 RMB per evaluated clip, equivalent to approximately 48 RMB per hour.
To ensure data quality, we applied our predefined validation protocol to screen and verify the collected responses.
In addition, we conducted a qualitative coding analysis of the textual justifications, assigning attribution codes to each response. The coding themes and procedural details are described in \Cref{app:code}.
% Each participant listens to 10 audio samples, including 3 trap items (1 flawed synthetic sample and 2 human recordings). 
% Response validity was established by requiring correct identification of the flawed synthetic clip and correct classification of at least one of the two human recordings; data from participants who failed this check were excluded from further analysis. 
All judgments, justifications, and demographic details were logged anonymously, and the study adhered to the ethical guidelines of the crowdsourcing platform and the researchers' institution.

% \paragraph{Generalization to more TTS systems. }

% \section{Results}

\subsubsection{Statistical Significance Test for ATT's Human Evaluation Protocol Design}\label{sec:stats}

To ensure that our results are statistically robust for subsequent analysis, we first conducted a statistical significance test using a Bayesian Generalized Linear Mixed Model (GLMM) \cite{bolker2009generalized}.
The GLMM analysis demonstrated excellent convergence and stability when applied to the human evaluation data.
All Gelman-Rubin diagnostics ($\hat{R}$) for the parameters were $1.00$, well below the convergence threshold ($\hat{R} < 1.01$), and the effective sample sizes (ESS) were all well above the recommended standard (ESS > 400), indicating high inference precision and reliable posterior estimates.

\begin{wraptable}{r}{0.5\linewidth}
    % \small
\vspace{-12px}
    \setlength{\tabcolsep}{3.5pt} % 缩小列间距
    \centering
    \caption{\textbf{Posterior summary statistics from the GLMM}. Including posterior means, standard deviations (SD), 95\% highest density intervals (HDI).}\label{tab:stats}
       \resizebox{\linewidth}{!}{
\begin{tabular}{lcc}
\toprule
Models             & Posterior Mean(SD)   & $95\% \text{HDI}$ \\ \midrule
Seed-TTS           & 0.417 (0.011) & [0.398, 0.438]   \\
MiniMax-Speech  & 0.387 (0.011) & [0.368, 0.407]   \\
Step-Audio            & 0.286 (0.011) & [0.266, 0.307]   \\
Cosyvoice          & 0.234 (0.010) & [0.214, 0.254]   \\
GPT-4o             & 0.138 (0.011) & [0.118, 0.158]   \\
\bottomrule
\end{tabular}
}
\vspace{-10px}
\end{wraptable}
The fixed effects analysis indicates that the mean scores of all evaluated models were statistically significantly higher than the zero baseline (with $95\% \text{HDI}$ entirely above zero). 
Detailed results are provided in \Cref{tab:stats}. The findings indicate that the Seed-TTS and Minimax-Speech models significantly outperformed the GPT-4o and Cosyvoice models, while the Step-Audio model showed intermediate performance.

The random effects analysis reveals significant baseline differences across participants, with the estimated standard deviation of random intercepts being 0.234 ($95\% \text{HDI} = [0.222, 0.246]$), suggesting substantial individual variability in overall scoring tendencies. Additionally, there was a moderately positive correlation in repeated evaluations of the same model by individual raters (random slope standard deviation = 0.108, $95\% \text{HDI} = [0.100, 0.116]$), indicating stable preferences or biases in participants' judgments of specific models.

\subsubsection{Benchmarking via Human Evaluation}
% \todo{for figures: 1) blackbox and white box -> split is acceptable; 2) sub-dimension results; }

\begin{figure}
% \vspace{-10pt}
    \centering
    \subfigure[The overall performance of each TTS systems along with the split of black-/white-box.]{
    \includegraphics[height=0.24\linewidth]{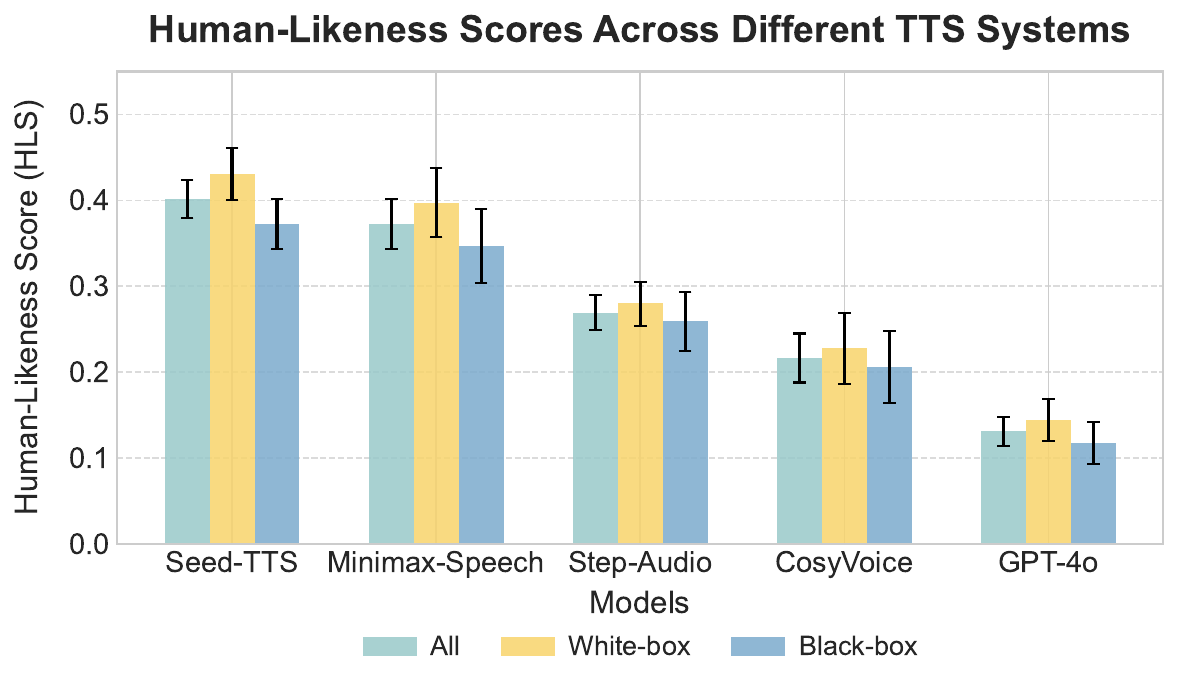}
    \label{fig:overall}
    }
    \hfill
    \subfigure[The TTS systems' performance in different capabilities dimensions.]{
    \includegraphics[height=0.24\linewidth]{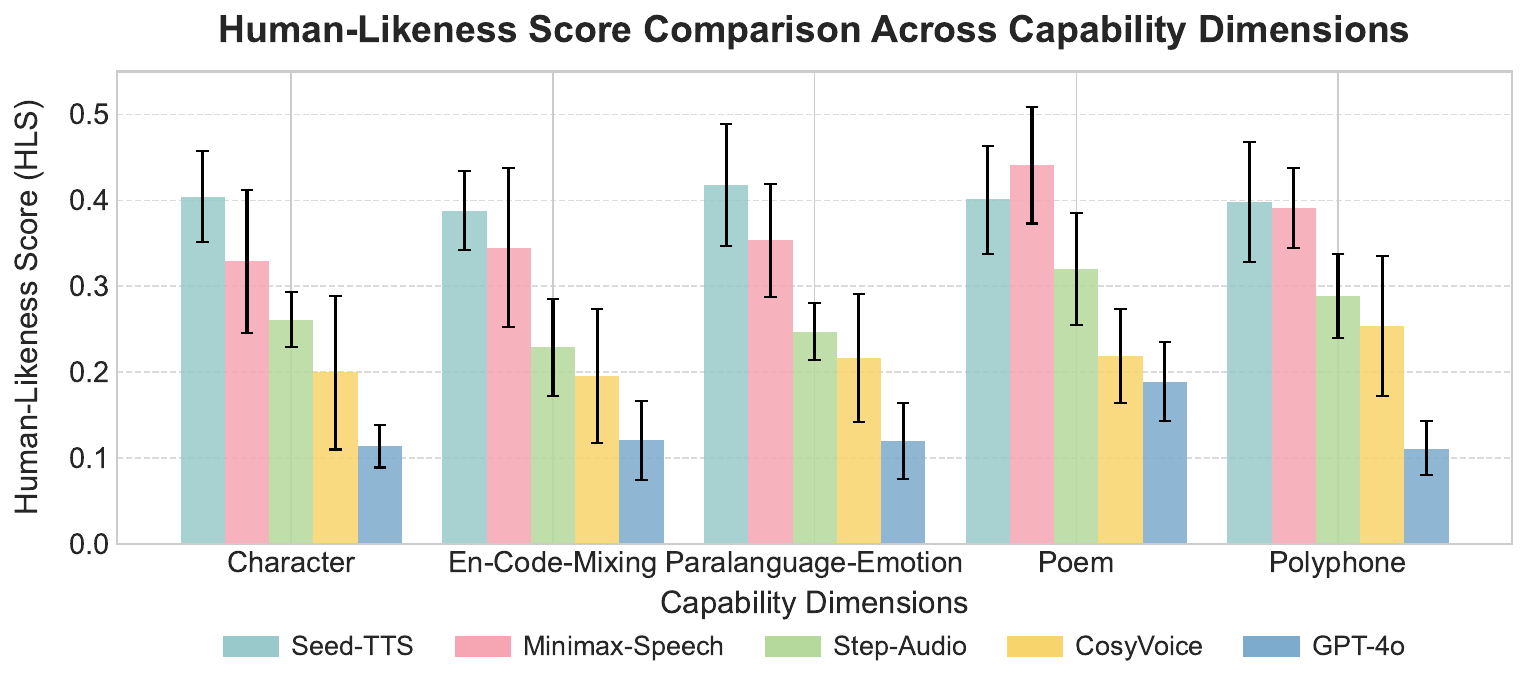}
    \label{fig:subdimension}
    }
    \vspace{-5pt}
    \caption{\textbf{The Key Benchmark Results of ATT Human Evaluation.} 
    }
    \vspace{-15pt}
\end{figure}

\paragraph{Effectiveness of ATT.} As shown in \Cref{fig:overall}, in ATT's benchmark results, Seed-TTS heads the first performance tier with Minimax-Speech. Step-Audio and Cosyvoice occupy the second tier with mid-range scores between 0.22 and 0.27, while GPT-4o falls into a distinct third tier at just 0.13, well below the leaders. 
The pronounced stepwise gaps show that the ATT evaluation framework can clearly distinguish capability differences among TTS systems.
\textbf{The most notable result is that the highest model's HLS is only 0.4 (Seed-TTS), which remains substantially below the level of true human-likeness.} 
This result markedly deviates from the MOS scores widely reported in prior studies, where TTS systems have often been rated as nearly indistinguishable from human speech.
\textbf{This discrepancy suggests that the HLS metric in the ATT framework is more sensitive and effective in capturing the subtle differences between synthetic and human speech, thereby providing a more realistic assessment of TTS human-likeness.}

% \notes{comparison with other metrics -> ATT is better in distinguish}
\paragraph{Soundness of the black-/white-box split.} Crucially, the overall performance hierarchy remains consistent when comparing white-box and black-box evaluation settings: each model retains the same relative ranking across both conditions (as shown in \Cref{fig:overall}).
The small and uniform performance gap between the two settings indicates that they are of comparable difficulty, confirming that the black-box/white-box split is well-balanced and does not introduce systematic bias into the evaluation.

\paragraph{Performance of Different Dimensions and Different Voice Styles.} 
Leveraging ATT's capability for cross-model comparison, we conducted a more fine-grained analysis of the human-likeness exhibited in different voice timbres generated by each TTS system, as well as their overall performance across multiple dimensions.
Importantly, as shown in \Cref{fig:subdimension} all the models' scores on each sub-dimension mirror their positions in the overall league table, showing no large fluctuations between individual skills and total capability. 
Notably, substantial variations in voice style are also observed within individual models.
For example, Seed-TTS's top-ranked voice, ``Skye,'' scores 0.47, whereas the lowest-ranked voice scores only 0.35. 
This clear gap shows that ATT can distinguish quality variations between different timbres generated by the same model.
The detailed results can be found in \Cref{app:detailresult}.

\paragraph{Attribution Analysis.} 
The qualitative review of the judges' comments reveals common shortcomings across all vendors: (1) prosodic naturalness: intonation patterns often appear abrupt or unnatural, with long sentences delivered in a word-by-word manner and lacking appropriate micro-pauses, making the synthetic origin readily detectable; and (2) expressive richness: emotional expression is either overly flattened or semantically incongruent with the content of the sentence. GPT-4o's Chinese voices are additionally hindered by a noticeable foreign accent, poor rhythm control, and prominent audio artifacts (electronic hiss and noise), which compound its prosodic issues and place it firmly at the bottom.

% \paragraph{Comparison with other Evaluation Methods}

\subsection{Effectiveness of \me Evaluation}\label{sec:modelexpsetup}

To validate the effectiveness of \me, we design experiments from two aspects: (1) comparing \me performance against other MOS-prediction models and (2) measuring \me alignment with human judgments.
\begin{figure}
    % \vspace{-15pt}
    \centering
    \includegraphics[width=\linewidth]{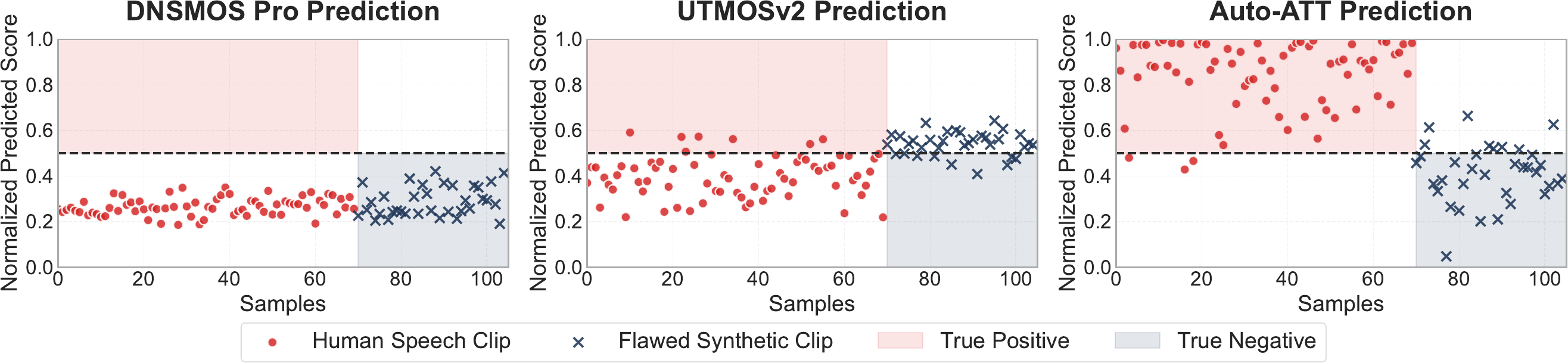}
    \caption{\textbf{The prediction results of Trap Item through DMSMOS Pro, UTMOSv2, and \me.} For a human speech clip, the ideal outcome is a true positive: the red dot should fall within the red zone; for a flawed synthetic speech clip, the ideal outcome is a true negative: the gray dot should fall within the gray zone. }
    \label{fig:trap}
    \vspace{-15pt}
\end{figure}

\subsubsection{Comparison with Other Auto Evaluation in Trap Item} 
To evaluate model reliability, we conduct experiments on the trap items included in the \ac. 
We compare the state-of-the-art automatic evaluation methods UTMOSv2 \cite{Baba2024UTMOSv2} and DMSMOS Pro \cite{cumlin24_interspeech} with our \me in predicted HLS on these trap items. Since trap items are readily distinguishable to human listeners in our data validation process, we scored them with each prediction model. 
In principle, a reliable model should accurately predict the quality of trap items. 
For both MOS prediction and ATT scores, human speech should receive significantly higher ratings than defective synthetic speech.
As shown in \Cref{fig:trap}, \me predicts trap items markedly better than conventional MOS prediction models. \me vastly outperformed the baselines, achieving an F1 score of 0.92, while UTMOSv2 reached only 0.14 and DNSMOSPro collapsed to 0.00 at the 0.5 decision threshold.
This result indicates that, in comparison to conventional MOS prediction models, \me demonstrates superior capability in distinguishing the human-likeness of speech audio, making it particularly well-suited for automated evaluation tasks.

\subsubsection{Consistency of Human Evaluation} 
\begin{wraptable}{r}{0.5\linewidth}
\vspace{-15pt}
\setlength{\tabcolsep}{3pt} % 缩小列间距
\centering
\caption{Kendall $\tau$ distance \textit{(p\,-value)} of \me\ and Qwen2-Audio-Instruct across different capability dimensions. Lower $\tau$ is better.}
\label{tab:kendall_compare}
\resizebox{\linewidth}{!}{
\begin{tabular}{lcc}
  \toprule
  \textbf{Capability Dimension} & \textbf{Auto-ATT} & \textbf{Qwen2 Audio} \\
  \midrule
    All                                          & \textbf{0.3316 (0.0398)} & 0.3474 (0.0638) \\
  \midrule
  \multicolumn{3}{c}{\textit{In-Distribution Dimensions}} \\
  \midrule
   Special Characters and Numerals              & \textbf{0.2737 (0.0047)} & 0.3105 (0.0198) \\
  Chinese--English Code-switching              & \textbf{0.3368 (0.0468)} & 0.3737 (0.1284) \\
  Paralinguistic Features and Emotions         & \textbf{0.2789 (0.0059)} & \textbf{0.2789 (0.0059)} \\
  \midrule
\multicolumn{3}{c}{\textit{Out-of-Distribution Dimensions}} \\ 
\midrule
  Classical Chinese Poetry/Prose               & \textbf{0.3421 (0.0548)} & 0.3526 (0.0740) \\
  Polyphonic Characters                        & \textbf{0.3316 (0.0398)} & 0.3474 (0.0638) \\
  \bottomrule
\end{tabular}}
\vspace{-15pt}
\end{wraptable}
To validate the alignment between \me predictions and human assessments, we used four NVIDIA A100 GPUs to run \me and Qwen2-Audio-Instruct on the same audio data as the human evaluation to predict their HLS scores. 
Each model took about two hours to process. 
We ranked each voice's human-likeness based on predicted HLS across each dimension, comparing these rankings to those derived from human evaluations using Kendall's distance \cite{abdi2007kendall}. 
As shown in \Cref{tab:kendall_compare}, \me consistently maintains rankings highly aligned with human judgments both overall and across individual dimensions. Furthermore, \me demonstrates superior consistency and assessment accuracy compared to Qwen2-Audio-Instruct, underscoring its enhanced capability for reliable human-likeness evaluation.

% \subsection{Effectiveness of \me}

% \paragraph{Align with human evaluation.} 

% \paragraph{Generalization of evaluating TTS systems.}

% \begin{wraptable}{r}{0.5\linewidth}
% \setlength{\tabcolsep}{3.5pt} % 缩小列间距
%     \centering
%   \caption{Kendall $\tau$ distance of \me and Qwen2-Audio-Instruct across different capabilities dimension. Lower is better.}
%   \label{tab:kendall_compare}
%   \resizebox{\linewidth}{!}{
%   \begin{tabular}{lcc}
%     \toprule
%     \textbf{Capabilities Dimension} & \textbf{Auto-ATT} & \textbf{Qwen2 Audio} \\
%     \midrule
%     All                    & 0.2737 & \textbf{0.3105} \\
%     Special Characters and Numerals                      & 0.2737 & \textbf{0.3105} \\
%      Chinese-English Code-switching   & 0.3368 & \textbf{0.3737} \\
%     Paralinguistic Features and Emotions  & 0.2789 & 0.2789 \\
%     Classical Chinese Poetry/Prose                          & 0.3421 & \textbf{0.3526} \\
%     Polyphonic Characters                     & 0.3316 & \textbf{0.3474} \\
%     \bottomrule
%   \end{tabular}
%   }
%   \vspace{-15pt}
% \end{wraptable}

% \section{Discussion}
\vspace{-2pt}
\section{Conclusion \& Limitations}\label{sec:limitations}
\vspace{-2pt}
In this paper, we propose the Audio Turing Test (ATT), an innovative evaluation framework specifically designed to address critical challenges in evaluating the human-likeness of LLM-based TTS systems in Chinese. 
ATT uniquely integrates a comprehensive, multi-dimensional evaluation corpus \ac with a robust Turing-Test-inspired evaluation protocol, thereby providing both qualitative and quantitative insights. 
Our rigorous validation demonstrates that ATT reliably differentiates among state-of-the-art LLM-based TTS models, pinpointing specific strengths and weaknesses across diverse linguistic dimensions such as code-switching, emotional expression, polyphony, and classical texts.
Additionally, by finetuning Qwen2-Audio-Instruct on human annotations, we develop \me for accelerating the iteration cycles of TTS systems through rapid and accurate assessments. 
Results confirm \me's superior alignment with human evaluators compared to traditional automatic evaluation methodologies.
A current limitation of ATT is its language-specific nature, as both the protocol and corpus are primarily designed for Chinese speech synthesis. To address this, we aim to extend the ATT framework to support multiple languages and a broader range of speech synthesis scenarios, thereby validating its generalizability and cross-linguistic effectiveness. Overall, ATT represents a significant advancement in the evaluation of large language model (LLM)-based speech synthesis systems. It paves the way for more natural and human-like text-to-speech technologies, fostering ongoing innovation and accelerating practical applications that enhance user experience.

% \clearpage

% \todo{limitations}
% \begin{ack}
% Use unnumbered first level headings for the acknowledgments. All acknowledgments
% More information about this disclosure can be found at: \url{https://neurips.cc/Conferences/2024/PaperInformation/FundingDisclosure}.

% Do {\bf not} include this section in the anonymized submission, only in the final paper. You can use the \texttt{ack} environment provided in the style file to automatically hide this section in the anonymized submission.
% \end{ack}

% \begin{ack}
% Use unnumbered first level headings for the acknowledgments. All acknowledgments

% Do {\bf not} include this section in the anonymized submission, only in the final paper. You can use the \texttt{ack} environment provided in the style file to automatically hide this section in the anonymized submission.
% \end{ack}
% \clearpage
\bibliography{att}
\bibliographystyle{plainnat}

%%%%%%%%%%%%%%%%%%%%%%%%%%%%%%%%%%%%%%%%%%%%%%%%%%%%%%%%%%%%
\clearpage
\appendix

% \section{Technical Appendices and Supplementary Material}
% Technical appendices with additional results, figures, graphs and proofs may be submitted with the paper submission before the full submission deadline (see above), or as a separate PDF in the ZIP file below before the supplementary material deadline. There is no page limit for the technical appendices.

\section{\ac Details}

\subsection{Data Generation}\label{app:ATTdata}
We found that we could not directly synthesize colloquial texts that met our requirements, so we designed a three-step corpus-creation workflow: 1) use GPT-4o \cite{hurst2024gpt} to batch-generate Chinese sentences that mix in English, 2) pass these sentences through DeepSeek-R1 \cite{guo2025deepseek} for a colloquial adapt, 3) have linguistics experts further enrich and diversify the text through rewriting and perform final verification.

\paragraph{Batch Generate.} We first employed GPT-4o \cite{hurst2024gpt} to generate texts tailored to each predefined capability dimension. For example, for the Chinese-English code-switching dimension, we began by using the following prompt to produce Chinese sentences that incorporate English words.
\begin{lstlisting}
\end{lstlisting}
\begin{CJK}{UTF8}{gkai}
给我一些日常沟通的的中文长文本，每一句话中需要有非常自然的中英文掺杂的现象，一句话只出现1-2个单词，且主要为专有词汇，或英文的filler words。

示例一：今天在朋友圈看到朋友发的自拍，她在用一个叫FaceTune的app修图，效果真的是很棒，很自然，你要不要也试试？

示例二：昨晚在Hulu上看了一部新的浪漫喜剧，叫《To All the Boys I've Loved Before》，剧情特别甜，看完之后觉得心情特别好。

示例三：今天在星巴克点了一杯新的Cold Brew Coffee，味道特别醇厚，喝完感觉一整天都特别清醒，推荐你也试试，很提神哦！

示例四：最近我一直在用 Estée Lauder 的粉底液，它的妆效很 natural，能够很好地贴合肌肤，遮盖瑕疵的同时又不会显得很厚重，让我的肌肤看起来自然无瑕，仿佛天生丽质一般。

示例五：今天我在网易云音乐上闲逛的时候，发现了一首超好听的新歌，叫《Shape of You》。那旋律可动感了，我听完之后，心情瞬间变得超好，感觉整个人都跟着节奏摇摆起来了，你听过这首歌吗？

执行后将每句话的长度拓展到100字左右。执行后将部分句子的句末加一些语气助词，丰富句子的口语化程度，但不要夸张。需要熟悉中国人的口语习惯，然后生成以上要求内容。请给我40句
\end{CJK}
\begin{lstlisting}
\end{lstlisting}

\paragraph{Colloquial Rewrite.} To make the text still more conversational, we ran it through DeepSeek-R1 \cite{guo2025deepseek} for an additional colloquial rewrite, using the prompt shown below:
\begin{lstlisting}
\end{lstlisting}
\begin{CJK}{UTF8}{gkai}
将给出的文本改写为更加口语化，有沟通感的文本，并添加一定的背景及前后连贯信息，你可以从以下的6个示例中获得灵感，但不允许照搬照抄，或者仿照句式，不允许用同样重复的开头

示例1：
原始：开始用Notion这个app之后，发现它真的太强大了，不仅可以用来记笔记，还能用来管理项目和计划，非常实用，简直是提高效率的利器呢。
更改为：我跟你讲，我最近在用Notion这个app，我的天我真的发现它真的很强，不仅可以用来记笔记，还能用来管理项目和计划，而且还很美观，真的挺实用的，是个提高效率的好东西，你们要不要也用一下看看？

示例2：
原始：朋友推荐了一个新的K-pop组合，叫BTS，听了他们的几首歌后真的觉得很好听，特别是那首《Dynamite》，旋律超级洗脑，推荐你也去听听看。
更改为：昨天跟家里那帮朋友出去吃饭，他们给我推荐了一个新出来的的K-pop组合，叫BTS，还挺不错，听了他们的几首歌都还可以，特别《Dynamite》这首，旋律超级洗脑，我从吃饭一直哼到回家洗澡，睡觉的时候脑子里都还在放这首歌，没救了。

示例3：
原始：下载了Pocket这个app，用来保存平时看到的好文章，觉得特别方便，这样有时间的时候就可以慢慢看，不会错过任何好内容，真的是读书神器。
更改为：天，哥们儿我跟你说，昨天我刚下载了Pocket这个app，发现它可以把平时看到的好文章都保存下来，也太方便了吧！你要不也用用看？这样有时间的时候就可以慢慢看，就不用担心错过很多不错的内容啦，真是读书神器绝绝子，安利你！

示例4：
原始：最近迷上了刷TikTok，真的有好多搞笑的短视频，看得我笑到不行，特别是那些创意短视频，简直让人一刷就停不下来，你也常常刷TikTok吗？
更改为：哇塞真的，TikTok一刷就停不下来，真的好多视频贼搞笑，短小精悍，看得我笑到不行！发明这些创意视频的博主也太有才了吧，好多时间一看就一两个小时过去了，你也刷TikTok吗？咱加个好友不。

示例5：
原始：昨晚在Hulu上看了一部新电影，叫《寄生虫》，剧情超精彩，每个情节都有出人意料的反转，看得我完全停不下来，一口气看完了整部电影，特别推荐。
更改为：你有看过最近大火的新电影《寄生虫》吗？我昨天在Hulu上看的，剧情好精彩啊，每个情节的反转都特别出人意料，根本想不到接下来会发生什么。其实我随手点开的，没想到会越看越上瘾，完全停不下来，最后一口气看完了，我跟你说你一定要去看，看完了记得和我分享。

示例6：
原始：开始用Headspace做冥想，每天花十分钟，整体状态变好了很多，特别是它的音指导很温柔，特别容易进入冥想状态，感觉整个人都特别放松。
更改为：最近不知道怎么了精神状态很差，所以我跟着一个叫Headspace的节目做冥想，每天花十分钟放空自己，练了快一个月，感觉自己压力没那么大了，睡眠质量也更好了，说起来我觉得这个channel最棒的是声音指导很温柔，你听了那个声音就很容易进入冥想状态，就觉得整个人好像在泡澡一样，特别安稳。
\end{CJK}
\begin{lstlisting}
\end{lstlisting}

\subsection{Audio Data Validation Criteria}\label{app:attaudio}
\textbf{Synthesis Success. }Synthesis success refers to the correctness of the output audio in terms of overall audio quality, transcription accuracy, and language appropriateness. 
Specifically, we check for issues such as: significant audio quality defects (e.g., excessive robotic noise, jittering), extremely short or incomplete audio that cannot be properly transcribed (e.g., only a single ``ah'' sound or complete silence), language mismatch (e.g., input in Chinese but output in Japanese), inconsistencies in voice timbre within a single clip (e.g., mixing multiple voice styles), and other cases where the output is unintelligible in the target language.

\textbf{Synthesis Consistency. }Synthesis consistency refers to the consistency of output when the same text is synthesized multiple times using the same voice and technology. This assessment focuses on whether the resulting audio clips are consistent in overall characteristics such as voice timbre (e.g., gender, age), language (e.g., remaining within the same language such as Chinese or English), and prosody (e.g., intonation, stress, and tone of voice). The goal is to determine whether the outputs can reliably be attributed to the same voice.

\section{ATT Benchmark Details}

\subsection{Evaluated TTS Systems}\label{app:tts}

Seed-TTS \cite{anastassiou2024seed} is ByteDance's large-scale foundation family for speech generation-its flagship autoregressive language-model variant scales into the multi-billion-parameter range and is trained with data and model sizes ``orders of magnitude larger'' than previous TTS systems, plus an optional diffusion decoder Seed-TTS-DiT.
Seed-TTS offers zero-shot speaker cloning, fine-grained emotion control and in-context speech editing while matching human naturalness scores in CMOS.

MiniMax-Speech-01 \cite{minimax_speech01} is an autoregressive Transformer TTS with an integrated learnable speaker encoder that enables true zero-shot voice cloning across 32 languages.
Although its exact size is undisclosed, the model is built on the same infrastructure as MiniMax-Text-01 (456B total/45.9B active parameters), so it inherits Mixture-of-Experts efficiency and ultra-long-context techniques from that 456B-parameter backbone.

CosyVoice2.0 \cite{du2024cosyvoice} delivers sub-150 ms first-packet latency in both streaming and offline modes, with multilingual zero-shot voice cloning across Chinese, English, Japanese, Korean and many dialects.
Public checkpoints of CosyVoice2.0 range from 300 M to 0.5 B parameters.

Step-Audio \cite{huang2025step} pairs a 130 B-parameter multimodal generative engine that synthetically bootstraps training data with a resource-efficient 3 B-parameter Step-Audio-TTS synthesiser.
This combination supports controllable speech with emotions, dialects and styles, and meets real-time requirements through speculative decoding and a dual-codebook tokenizer architecture.

OpenAI's GPT-4o \cite{hurst2024gpt} is an end-to-end multimodal model (parameter count not publicly disclosed) that handles text, vision and audio in one network and speaks with human-like latency-232 ms best-case, 320 ms on average.
It matches GPT-4-Turbo on text but adds expressive speech synthesis, real-time translation and paralinguistic cues without the separate ASR and TTS stages used in previous Voice Mode pipelines.

% \subsection{Participants}
%

\subsection{Instructions and User Interface}\label{app:ui}

We provide instructions for each participant for the evaluation task and design the reward system to encourage the high-quality evaluation. 

Since our benchmark are in Chinese, our instructions are also in Chinese for native speaker participants.
Here we provide a translated English version for review:
\begin{lstlisting}
Task description

In this task, you must decide whether each audio clip you hear is spoken by a real person or generated by a machine, and you must state why you reached that conclusion.
Your written reason is the main evidence used in manual review, so base it on concrete observations of the recording.

For every 10 clips there are several hidden "test items."
These have an unmistakably correct answer; selecting the wrong answer on a test item will cause your entire submission to fail review. Do not rely on AI to draft your responses-judgements that fail review will be discarded and not counted as valid data.


How to write your reason

Examples of poor reasons

(Not convincing; give no specific evidence from the audio)
	1.	"Pure machine voice."
	2.	"The imitation of human speech is too forced."
	3.	"Obviously a machine tone-doesn't sound like a real person."
	4.	"Sounds like a late-night radio host."

Examples of good reasons

(Accurate analysis that cites concrete details in the clip)
	1.	The phrase "Many thins" should end with a falling intonation, but here it rises-it sounds unnatural.
	2.	The clip is machine-generated: each word pops out individually with poor flow.
	3.	The phrase "go away" lacks the angry/impatient tone that should be present.
	4.	After the word "angry," the breath has a noticeable electronic/robotic artifact.
\end{lstlisting}

\begin{figure}[H]
    \centering
    \includegraphics[width=0.9\linewidth]{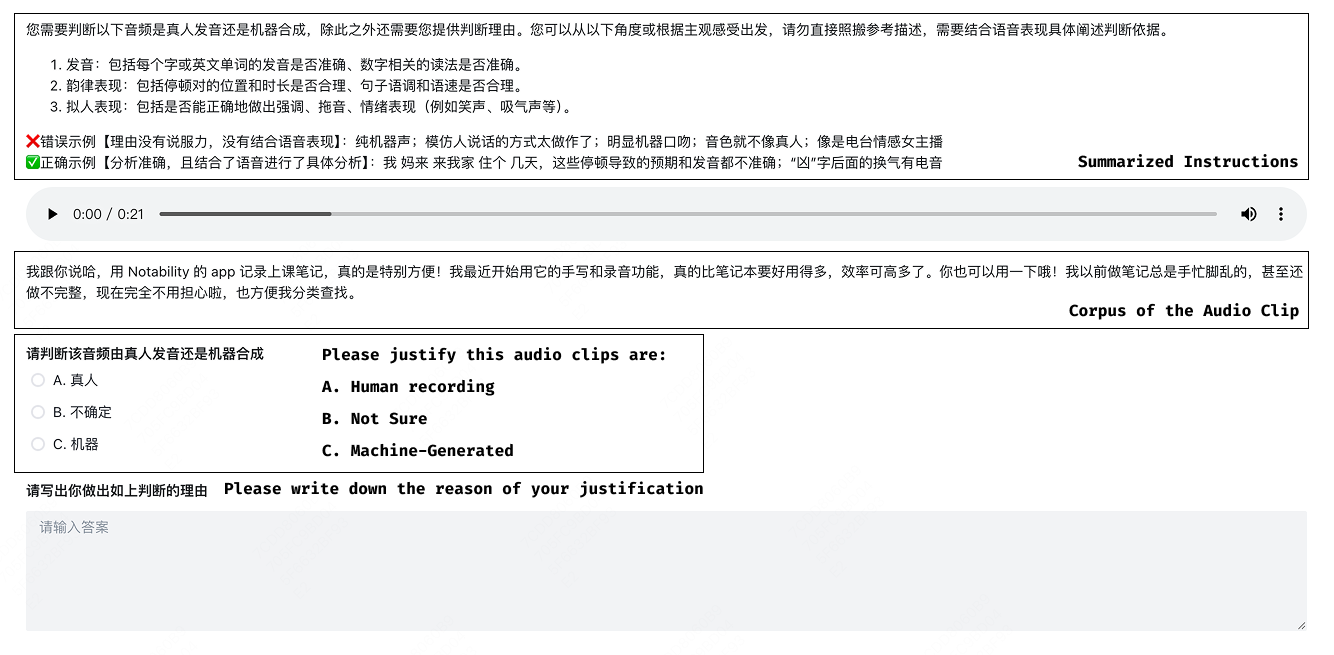}
    \caption{The Screen of One Audio Clip in ATT Evaluation.}
    \label{fig:ui}
\end{figure}

And the user interface for the task are shown in \Cref{fig:ui} with explanation in English.

\subsection{Qualitative Analysis}\label{app:code}

The coding criteria for qualitative analysis are based on \Cref{tab:qualitative_analysis}, which consists of four dimensions: first, pronunciation accuracy, focusing on the correctness of each Chinese character's pronunciation (especially polyphonic characters within words), accuracy of tones, correct pronunciation of embedded English words, and accurate pronunciation of numerical information such as dates, monetary amounts, and phone numbers; second, prosodic appropriateness, examining whether pauses occur at reasonable positions with appropriate duration, whether the sentence intonation matches semantic intentions (e.g., questions or exclamations), and whether speech speed is appropriate without being excessively fast or slow; third, audio clarity, assessing overall audio quality, including the presence of noticeable background noise, jitter, or electronic distortion in pronunciations; and fourth, naturalness and human-like expressiveness, evaluating whether the overall speech performance appears human-like and natural by considering factors such as semantic emphasis and prolongation of words, emotional expressions consistent with sentence meaning, and effective paralinguistic features including breaths, laughter, crying, coughing, or breathy voice.

\begin{table}[htbp]
\centering
\caption{Criteria for Qualitative Analysis}
\begin{tabular}{p{4cm}p{9cm}}
\toprule
\textbf{Dimension} & \textbf{Detailed Explanation} \\ \midrule
Pronunciation Accuracy &

     - Whether each Chinese character is pronounced correctly, especially polyphonic characters within words.
     - Whether the tones of characters/words are accurate.
     - Whether embedded English words are pronounced correctly.
     - Whether numerical information such as dates, monetary amounts, and phone numbers is read accurately.

\\ 
Prosodic Appropriateness &

     - Whether the position and duration of pauses are reasonable.
     - Whether the intonation matches the sentence meaning, such as questions or exclamations.
     - Whether speech speed is appropriate, avoiding overly fast or slow pacing.

\\ 
Audio Clarity &

     - Whether the overall audio quality is clear, and if noticeable background noise is present.
     - Whether pronunciations have jitter, electronic distortion, or other clarity issues.

\\ 
Naturalness and Human-like Expressiveness &
     - Whether the overall speech appears natural and comparable to human speech, considering:
    \begin{itemize}
        \item Appropriate semantic emphasis on words.
        \item Appropriate prolongation of words matching semantic context.
        \item Emotional expressions matching the sentence context.
        \item Effective use of paralinguistic features such as breathing sounds, laughter, crying, coughing, or breathy voice.
    \end{itemize}
\\ \bottomrule
\end{tabular}
\label{tab:qualitative_analysis}
\end{table}

\subsection{Detail Results}\label{app:detailresult}

\paragraph{Dimensional Performance.}
Across ATT's five evaluation dimensions, Seed-TTS consistently ranks first, demonstrating the strongest overall performance and particularly excelling at Chinese-English Code-switching and Special Characters and Numerals; its only relative weakness is in Classical Chinese
Poetry/Prose, where it is narrowly outperformed by Minimax-Speech. 
Step-Audio, Cosyvoice, and GPT-4o follow in that order. 

\paragraph{Different Voice Styles Performance.} 
We list the performance of each voice style in \Cref{tab:vs_details}.

\begin{sidewaystable}
\centering
\caption{HLS of Different Voice Styles}
\label{tab:vs_details}
\resizebox{0.8\paperheight}{!}{
\begin{tabular}{ccccccc}
\toprule
\textbf{Model} & \textbf{Voice Style} & \textbf{Special Characters and Numerals} & \textbf{Chinese-English Code-switching} & \textbf{Paralinguistic Features and Emotions} & \textbf{Classical Chinese Poetry/Prose} & \textbf{Polyphonic Characters} \\
\midrule
\multirow{4}{*}{CosyVoice}
  & longshuo       & 0.1075 & 0.1175 & 0.135 & 0.18 & 0.17 \\
  & longxiaocheng  & 0.211 & 0.1275 & 0.14 & 0.1875 & 0.1975 \\
  & longxiaochun   & 0.125 & 0.2325 & 0.2615 & 0.2215 & 0.2625 \\
  & longxiaoxia    & 0.355 & 0.305 & 0.33 & 0.285 & 0.385 \\
\hline
\multirow{4}{*}{MiniMax-Speech}
  & siyuan         & 0.2775 & 0.3125 & 0.3075 & 0.365 & 0.329 \\
  & xinyue         & 0.4575 & 0.417 & 0.45 & 0.515 & 0.4325 \\
  & yaoyao         & 0.3625 & 0.4275 & 0.35 & 0.455 & 0.4275 \\
  & zixuan         & 0.2175 & 0.225 & 0.3075 & 0.43 & 0.375 \\
\hline
\multirow{4}{*}{Seed-TTS}
  & Alvin          & 0.4 & 0.36 & 0.395 & 0.4 & 0.3625 \\
  & Brayan         & 0.4125 & 0.3925 & 0.36 & 0.405 & 0.4295 \\
  & moon           & 0.365 & 0.36 & 0.4 & 0.3225 & 0.3 \\
  & skye           & 0.44 & 0.44 & 0.5175 & 0.475 & 0.5 \\
\hline
\multirow{4}{*}{GPT-4o}
  & alloy          & 0.1525 & 0.095 & 0.155 & 0.171 & 0.1005 \\
  & echo           & 0.085 & 0.0975 & 0.0555 & 0.1425 & 0.075 \\
  & onyx           & 0.1 & 0.135 & 0.095 & 0.196 & 0.12 \\
  & shimmer        & 0.1175 & 0.155 & 0.175 & 0.246 & 0.15 \\
\hline
\multirow{4}{*}{Step-Audio}
  & wenjingxuejie        & 0.2325 & 0.252 & 0.2375 & 0.3625 & 0.329 \\
  & shenchennansheng     & 0.2425 & 0.214 & 0.2575 & 0.2425 & 0.283 \\
  & linjiajiejie         & 0.2655 & 0.181 & 0.21 & 0.27 & 0.2125 \\
  & qingniandaxuesheng   & 0.304 & 0.2675 & 0.285 & 0.405 & 0.3315 \\
\bottomrule
\end{tabular}
}

\end{sidewaystable}

\end{document}